\documentclass[useAMS,usenatbib]{mn2e}

\usepackage{graphics}
\usepackage{epsfig}

\title[NGC 1407 Globular Clusters]{Evidence for Two Phases of Galaxy Formation from 
Radial Trends in the Globular Cluster System of NGC 1407}
\author[D. A. Forbes et al.]{Duncan A. Forbes$^{1}$\thanks{E-mail:
dforbes@swin.edu.au}, Lee R. Spitler$^{1}$, 
Jay Strader$^{2}$
Aaron J. Romanowsky$^{3}$, 
\newauthor
Jean P. Brodie$^{3}$, Caroline Foster$^{1}$ 
\\
$^{1}$Centre for Astrophysics \& Supercomputing, Swinburne
University, 
Hawthorn VIC 3122, Australia\\
$^{2}$Harvard-Smithsonian Cetre for Astrophysics, Cambridge, MA 02138, USA\\
$^{3}$UCO/Lick Observatory, University of California Santa Cruz, CA 95064, USA
}
\begin{document}


\pagerange{\pageref{firstpage}--\pageref{lastpage}} \pubyear{2002}

\maketitle

\label{firstpage}

\begin{abstract}

Here we present the colours of individual globular clusters (GCs)
around the massive elliptical galaxy NGC 1407 out to a projected 
galactocentric radius of 140 kpc or 17 galaxy effective radii (R$_e$). Such
data are a proxy for the halo metallicity. We find steep, and
similar, metallicity gradients of $\sim$ --0.4 dex per dex 
for both the blue (metal-poor) and
red (metal-rich) GC subpopulations within 5--8.5 R$_e$ (40--70
kpc). At larger
radii the mean GC colours (metallicity) are constant. A similar
behaviour is seen in a wide-field study of M87's GC system, and
in our own Galaxy. We interpret these radial metallicity trends
to indicate an inner region formed by early {\it in-situ}
dissipative processes and an outer halo formed by ongoing
accretion of low mass galaxies and their GCs. These results provide
observational support for the model of galaxy formation whereby
massive galaxies form inside-out in two phases. We have also
searched the literature for other massive early-type galaxies
with reported GC metallicity gradients in their inner regions. No
obvious correlation with galaxy mass or environment is found but
the sample is currently small.

\end{abstract}


\section{Introduction}

An emerging picture of galaxy formation is that massive early-type
galaxies grow from relatively compact `seed' galaxies with
effective radii of $\sim$ 1 kpc at z $\ge$ 2 (sometimes referred to as `red
nuggets') to today's massive systems of effective radii of $\sim$
5 kpc (Oser et al. 2010; Gobat et al. 2010). 
This size evolution has a redshift
dependence of the form $(1+z)^{-1.2}$ (Franx et al. 2008;
Williams et al. 2010) and is driven by the growth of the outer
regions of galaxies that experience ongoing accretion (van Dokkum et
al. 2010). 

According to theory, 
the growth of stellar mass occurs in two phases: 1) An {\it
in-situ} dissipative phase with cold flows (Keres et al. 2005;
Dekel et al. 2009; Naab et al. 2009; Oser et al. 2010) that is reminiscent of a
`monolithic collapse' (Larsen 1975) and 2) ongoing accretion from
minor mergers of lower mass systems (Abadi et al. 2006; 
Hopkins et al. 2009; Zolotov
et al. 2010). The {\it in-situ} stars form, and are in place,
early in the Universe (i.e. $\ge$ 10 Gyrs ago), whereas the outer
halo experiences a more continous growth as additional material
is accreted over time. 
The transition between the {\it in-situ} dominated inner regions
and the accretion dominated outer halo is predicted to occur 
at radii of tens of 
kiloparsecs (Abadi et al. 2006; Kolotov et al. 2010), with more
massive galaxies undergoing more accretion and having larger
transition radii (Oser et al. 2010).

This theoretical picture is difficult to test observationally.
For very nearby ($\sim$ 10 Mpc) early-type galaxies,
individual stars can be resolved by the {\it Hubble Space
Telescope} (HST) and the radial metallicity distribution probed. 
For example, 
in NGC 3379 the halo shows a metallicity transition at 
10--13 R$_{e}$, becoming metal-poor (Harris et al. 2007). Only a
handful of massive galaxies can be studied in this way. 

Intergrated light observations are required to obtain the
metallicity distribution for more distant galaxies. However, 
the surface brightness profiles of early-type galaxies fall off
rapidly with galactocentric radius making it difficult to probe
beyond a few effective radii. An example is the 
recent work of Coccato et al. (2010) who, after 13.5
hrs integration on an 8m telescope, 
obtained stellar population gradients in the massive 
Coma elliptical galaxy NGC 4889 to an unprecedented 4 
R$_{e}$. 
They derived metallicity, alpha element and age gradients from an 
analysis of the absorption lines using Lick indices. They found a
``break'' radius at $\sim$1.2R$_e$ (18 kpc) dividing the inner galaxy that
exhibited a steep metallicity gradient and 
high alpha element ratio (indicative of rapid star formation) from the
outer region that had a metallicity gradient consistent with zero
and evidence for more prolonged star formation. They described
the inner region as formed in an early dissipative collapse and
the outer region as having an accreted origin. Given the
telescope time involved it will be 
difficult to repeat such a study for a large sample of ellipticals. 

Another tracer of halo stellar populations is therefore required. Globular
clusters (GCs) have many advantages in this regard. As some GCs 
are destroyed over time, they contribute directly to the
halo field star population. In the case of the Milky Way, about
1/3 of the original GC system may have been destroyed 
(Mackey \& van den Bergh 2005) making up half of 
the current halo stellar mass (Martell \& Grebel
2010). When a low mass galaxy is accreted by a larger one, GCs
can survive the accretion process (Forbes \& Bridges 2010)
and the properties of these GCs contain 
information about their original host galaxy.   
Also, GC systems can extend beyond 100 kpc and hence they probe galaxy halos
to many effective radii. 

The globular cluster systems of large galaxies are generally found
to consist of two subpopulations -- blue (or metal-poor) and red
(or metal-rich). These are thought to be associated with the halo
and bulge/spheroid component of galaxies, respectively (Brodie \& Strader
2006). 
The mean colour of both subpopulations have been
shown to correlate with the luminosity or mass of the host
galaxy, with some possible flattening of the relation for low
mass galaxies 
(Forbes, Brodie \& Grillmair 1997; Larsen et al. 2001; Strader,
Brodie \& Forbes 2004; Peng et
al. 2006). We note that the mean colour used in such studies is
generally derived from the central regions of a galaxy.
Recent studies of extragalactic GC systems have revealed two
additional trends: the colour (metallicity) of GCs with
their magnitude (mass) and colour with projected galactocentric radius.  

The first trend, called the 
`blue tilt' (a tendency for the blue GCs to 
have redder colours at brighter magnitudes), was first discovered using the 
Advanced Camera for Surveys (ACS) on the 
HST to examine extragalactic GC systems. The ACS
provides highly accurate photometry and 
GC candidate lists that are largely free of contaminants (due to
the fact that GCs are partially resolved out to 
distances of $\sim$ 20 Mpc). The blue tilt has now been reported in a 
variety of galaxies including the most massive ellipticals (Harris et
al. 2006; Harris 2009a), 
lower mass ellipticals (Strader et al. 2006), early-type
dwarfs (Mieske et al. 2006a), early-type spirals (Spitler et
al. 2006) and a Milky Way like spiral (Forbes et al. 2010). 
It has not been detected in the Milky Way GC system (most likely due to
the small number of massive GCs) nor the massive elliptical NGC
4472 (Strader et
al. 2006; Mieske et al. 2006a) despite its large GC system.

Given the evidence that extragalactic GCs are mostly old (Brodie \& Strader 
2006), the blue tilt implies a mass-metallicity relation for the metal-poor 
subpopulation of GCs.
Various explanations have been proposed 
(see Bekki et al. 2007; Mieske et al. 2006a; Mieske 2009), but 
perhaps the most plausible explanation to date is one of 
self enrichment, whereby more massive GCs are enriched with 
heavier metals during their brief formation period (Parmentier
2004; Strader \& Smith 2008; Bailin \&
Harris 2009).


The second trend, whereby the mean colour of each GC subpopulation
varies with galactocentric radius, was first seen in ground-based
photometry of M49 using metallicity-sensitive filters (Geisler et
al. 1996). Such a trend needs to be carefully separated from the
trend of a declining mean colour with radius in the
{\it overall} GC system (due to the changing relative proportions of
blue and red subpopulations), and has only been measured for a handful of
other systems (Forte et al. 2001; Bassino et al. 2006; 
Harris 2009a,b). These studies 
indicate a negative metallicity gradient 
for both the blue and red subpopulations in
elliptical galaxies. 
In the case of the Milky Way, a metallicity
gradient is seen in both GC subpopulations within the central 10
kpc of slope $\approx$ --0.30 dex per dex (Harris 2001). 
Beyond
10 kpc, the mean GC metallicity in the Galactic halo is constant at 
[Fe/H] $\sim$ --1.5.
The M31 GC system reveals a strong metallicity gradient within 50
kpc, but flattens out to a near constant metallicity of [Fe/H]
$\sim$ --1.6 (Alves-Brito et al. 2009). 
These strong metallicity gradients in the inner halo regions
suggest a 
dissipative formation process {\it for both GC subpopulations}. 

Here we present the radial distribution of GC colour (and hence
metallicity) using deep wide-field Subaru imaging of the massive
elliptical galaxy NGC 1407 out
to $\sim$140 kpc or 17 R$_e$.
We find a strong radial gradient which is supported by HST
imaging and Keck spectroscopy. 
We compare the radial gradient to those
of other massive ellipticals, finding that the gradient
in the NGC 1407 GC system is the steepest published to date.
We briefly discuss the
wider implications for galaxy formation from our results.

\section{NGC 1407} 

NGC 1407 is the central 
massive elliptical in a large group (Brough
et al. 2006) with a rich globular cluster system (Forbes et
al. 2006). The galaxy itself reveals a steep metallicity
gradient in the starlight (Spolaor et al. 2008; Foster et
al. 2009) and is uniformly old with a supersolar alpha element
ratio within one effective radius. Our previous optical spectra
of 20 bright GCs in NGC 1407 indicated that they are mostly old
($\sim$11 Gyr) with a range of metallicity --1.5 $<$ [Z/H] $<$
0.0 (Cenarro et al. 2007). Some of the spectra showed evidence for
the presence of blue horizontal branch stars.
Our previous dynamical study of the GC system
(Romanowsky et al. 2009) found weak evidence for rotation
and a bias towards tangential orbits in the outer regions. We derived a
total virial halo mass of 6 $\times$ 10$^{13}$ M$_{\odot}$ which
indicates that NGC 1407 and its surrounding group is extremely
dark (with M/L$_B$ $\sim$ 800).  We assume an effective radius
for NGC 1407 itself of 1.17 arcmin as adopted by Harris (2009) and
Foster et al. (2009).

Images of NGC 1407 have been taken using the ACS camera on board the
HST using F435W (B) and F814W (I) filters covering 3.5 arcmin
$\times$ 3.5 arcmin. An analysis of the GC system using these
data has been reported by Forbes et al. (2006) and Harris et
al. (2006) and Harris (2009a). 
The GC system reveals a classic blue/red bimodality with 
a blue tilt in the blue subpopulation.

Harris (2009a) studied the colour-radius distribution for the NGC 1407
GC system finding an insignificant (1$\sigma$) trend for both the
blue and red GCs examined. However, when the NGC 1407 GC
system was combined with that of 5 other massive elliptical
galaxies, a statistically significant metallicity gradient of --0.10 dex per
dex was seen at
$>$4.5$\sigma$ for both subpopulations. 

The distance to NGC 1407 is somewhat uncertain. 
Here we adopt the mean value of 15 distances given by NED,
i.e. m--M = 31.85 $\pm$ 0.37 (or 23.77 $\pm$ 3.8 Mpc). At this distance,
1 arcmin equals 6.9 kpc.

\section{Photometric Data}

Here we use wide-field (34 arcmin $\times$ 27 arcmin) g,r,i images taken
using the Suprime-Cam camera on the Subaru 8m telescope. Total exposure
times were 3hr, 1hr, 0.9hr for the g,r,i images respectively. The
seeing was good at around 0.5 arcsec in each filter. 
The Suprime-Cam has a pixel
scale of 0.2$^{''}$ per pixel. 
Full details of the observations, data reduction and analysis
will be presented in Spitler et al. (2011, in prep.). Here we
give a brief summary.

The images were reduced and combined using the Suprime-cam
standard pipeline software. The final mosaic was 
placed on the USNO-B2 astrometric system using common stars. 
Before the initial object detection, a smooth model
of NGC 1407 (and the galaxy NGC 1400, which lies at a projected
separation of 12 arcmin) was
subtracted, using the ELLIPSE task in IRAF to help with object detection. 
DAOFIND was then used to select objects with a
threshold $\sim$3 times the standard deviation of the background count
level. 

For each point-like object an aperture magnitude and position was measured.
Aperture magnitudes were corrected to total magnitudes assuming
that the objects were point sources (a reasonable assumption for
GCs with an average expected size of 3 pc or 0.03 arcsec under
0.5 arcsec seeing) and applying filter-dependent aperture corrections.  
These were converted into standard magnitudes using
zeropoints determined with standard star observations. 
Extinction corrections using the maps of
Schlegel et al. (1998) were applied assuming no variation across
the image as NGC 1407 is known to be relatively free of dust from
mid-IR observations (Temi et
al. 2009).
Hereafter we quote extinction corrected magnitudes and colours. 
A colour selection was applied so that objects within 1$\sigma$ of 
the GC sequence in g--r and r--i colour-colour parameter space 
(i.e. 0.4 $<$ g--i $<$ 1.4 and 0.3 $<$ g--r $<$ 0.9) were included.

Finally, any GC candidates associated within 6 arcmin radius 
of NGC 1400 were removed from the
object list for the subsequent analysis. 
This effectively removes any possible contamination
from the GC system of NGC 1400. 

\section{Results}

In Figure 1 we show the distribution of GC colour with
projected 
galactocentric radius out to 20 arcmin from our wide-field Subaru imaging.
As well as the individual data points we show the mean value and
error on the mean in several radial bins using the NMIX statistical
test (Richardson \& Green 1997). 
The figure shows only those GCs with i $<$ 24 (M$_i$ $>$
--7.9). 
The data show evidence for a
clear negative metallicity gradient out to $\sim$6--10 arcmin
(5--8.5R$_e$ or 40--70 kpc) and are consistent with zero gradient beyond
that. A simple linear fit to radii within 10 arcmin gives central
colours of g--i = 0.880 $\pm$ 0.011 and 1.186 $\pm$ 0.007 for the
blue and red subpopulations respectively. Similarly, the slopes
are found to be 
--0.0084 $\pm$ 0.022 and --0.013 $\pm$ 0.002 mag per arcmin. Thus the blue
subpopulation gradient is at the $\sim$4$\sigma$ level and the
red one at the $\sim$6$\sigma$.

Following the method of Harris (2009b) we also fit the entire colour
gradient with the form g--i = $a + b$~log(R/R$_e$), where R$_e$ =
1.17 arcmin. We fit only the data points interior to
$\sim$10 arcmin (although including radial bins beyond this radius
makes little difference to the final values). For the blue
subpopulation we find $a$ = 0.818 $\pm$ 0.01 and $b$ = --0.063
$\pm$ 0.01 mag per dex. Similarly for the red subpopulation we find $a$ =
1.12 $\pm$ 0.02 and $b$ = --0.068 $\pm$ 0.02 mag per dex. For both
subpopulations the uncertainty quoted comes from the range of fits that
include/exclude the innermost data point (the steeper slope being
associated with an exclusion of the innermost radial bin).
Within the combined errors, the blue and red GC subpopulations 
show a similar colour gradient.

To convert the g--i colour gradients into 
metallicity gradients we follow Harris (2009a), i.e.  
$\Delta$g--i = 0.21$\Delta$[Fe/H], giving 
metallicity gradients of 
--0.38 $\pm$ 0.06 dex per dex 
for the blues and --0.43 $\pm$ 0.07 dex per dex for the reds. 

We also show in Figure 1 metallicities corresponding to g--i
colours assuming the linear relation found by
Sinnott et al. (2010). This relation is in part based on Milky
Way GC metallicities from Harris (1996) which is close to the
Carretta \& Gratton (1997) scale and hence we assume is closer to
total metallicity [Z/H] (see also discussion by Mendel et
al. 2007). 

We have calculated the probability density function (PDF) for the
data shown in Figure 1 in six radial bins (centred on 0.92, 1.81, 2.72,
3.85, 5.23 and 7.04 arcmin) of equal GC number. The PDF is sensitive
to colour substructure and is shown graphically in
Figure 2. Other than the innermost radial bin, 
the PDF for the blue GCs confirms their radial colour gradient.
For the red GCs the radial trend is less clear due to
possible colour substructure in the inner regions. 

In Figure 3 we show the distribution of GC colour with
galactocentric radius from the HST ACS imaging (Forbes et
al. 2006). These data have the advantage of having a lower
contamination rate than the Subaru imaging (which we estimate to
be less than 5\% for bright GCs based on follow-up spectra;
Romanowsky et al. 2009) but probe to much smaller galactocentric
radii. Again we show the mean colours and the error on the mean
in various radial bins with equal numbers of GCs. A clear colour
gradient is seen in the red subpopulation and to a lesser extent
in the blue subpopulation supporting our Subaru results
above. Fitting B--I = $a + b$~log(R/R$_e$) we find $a$ = 1.62
$\pm$ 0.02 and $b$ = --0.147 $\pm$ 0.065 dex per dex for the blue
subpopulation and $a$ = 2.05 $\pm$ 0.01 and $b$ = --0.169 $\pm$
0.005 dex per dex for the red GCs.

Assuming a conversion of B--I colour
into metallicity of $\Delta$B--I = 0.375$\Delta$[Fe/H] (Harris
2009b), these slopes give a metallicity gradient of 
--0.39 $\pm$ 0.17 dex per dex 
for the blues and --0.45 $\pm$ 0.01 for the reds. These values 
are identical to those 
inferred from the larger radii g--i gradient.

We also show in Figure 3 the metallicities corresponding to B--I 
colours assuming the linear relation quoted by Harris (2009b).
As for the g--i colours, this relation based on Milky
Way GC metallicities from Harris (1996) which is close to the
Carretta \& Gratton (1997) scale and hence total metallicity [Z/H].

The radial stellar population properties for the NGC 1407 field
stars out to $\sim$0.5R$_e$ have been measured, from an analysis of
the absorption lines, by Spolaor et al. (2008). They found
the galaxy to have a central total metallicity of [Z/H] $\sim$ +0.3 dex
and a steep metallicity gradient of --0.38 $\pm$
0.04 dex per dex. These values are 
similiar to those found for the red GCs in NGC 1407 and are
illustrated in Figure 3. The error on individual points from
Spolaor et al. is $\pm$
0.15 dex. 
This supports
previous claims that the red GCs are closely associated with the
bulge/spheroid component of elliptical galaxies (Forbes \& Forte
2001; Forte et al. 2009; 
Spitler 2010). The metallicity gradient of Foster et al. (2009)
has a similar slope but has individual errors of $\pm$ 1 dex (and
is not shown).


Could the colour gradient for the blue subpopulation 
seen in Figures 1, 2 and 3 simply be due to a
changing strength of the blue tilt with radius? For this to be
the case the blue tilt would need to be significantly stronger (i.e. tilted
more to the red at bright magnitudes) in the inner galaxy regions
and to disappear (i.e. no blue tilt) for galactocentric radii
beyond $\sim$10 arcmin. Mieske et al. (2006, 2010) found 
weak evidence for a stronger blue tilt at
small galactocentric radii in their data for Virgo and Fornax early-type
galaxies. 
However, for the large galaxies in their study, they were
generally restricted to the inner $\sim$1.5R$_e$ due to the HST ACS
field-of-view. Harris (2009a) explored this issue out to $\sim$5R$_e$
using a composite of the GC systems in 6 giant ellipticals, 
finding no convincing evidence for a stronger blue tilt in the
inner regions. Further evidence comes from the fact that NGC 4472
reveals a radial gradient (Geisler et al. 1996) but no blue
tilt (Strader et al. 2006). 


We have divided the Subaru 
data into 3 radial bins, i.e. to a projected
galactocentric radius of 5 arcmin, between 5 arcmin and 10 arcmin
and beyond 10 arcmin. All three radial bins reveal a blue tilt of
similar strength. 
The relatively constant strength of the blue tilt with
radius can not explain the observed gradient in the mean colour
with radius within 10 arcmin (nor the lack of a gradient beyond
10 arcmin). We thus conclude that the colour gradient observed in
the blue subpopulation is not caused by a changing
colour-magnitude trend with radius. Finally, the radial
metallicity gradient of the blue subpopulation is similar to that
of the red GCs, which do not show a `red tilt'.

Could the GC colour gradients be due to contamination? 
The radial surface density profiles of the 
blue and red GC candidates around NGC 1407
(Spitler et al. 2011, in prep), show that they are dominated by
objects that are associated with NGC 1407 for radii $<$10 arcmin
(i.e. for which the strong metallicity gradient is seen). Based
on these profiles the estimated contamination rate at 10 arcmin
is less than 20\% for the blue GCs but could be as high as 50\%
for the red GCs. Thus the transition radius for the blue GCs
appears to be robust but the flattening of the gradient for the
red GCs at large radii could be an artifact of increased contamination.

Foster et al. (2010) explored the use of the Calcium Triplet (CaT)
lines to derive spectroscopic metallicities for GCs in NGC 1407
but did not explore the radial variation of CaT metallicity for
the GC subpopulations.
Using a linear CaT to metallicity transformation based on Milky Way GCs by
Armandroff \& Zinn (1988), they found the red GCs to have
reasonable metallicities but the blue GCs had inferred
metallicities that were too high, i.e. they had stronger CaT
strengths than expected. This could be due to an extra
contribution from blue horizontal branch stars (as suggested by
Cenarro et al. 2007 for the NGC 1407 GCs). In Figure 4 we plot
the CaT derived metallicities (converted into total metallicity
assuming [Z/H] = [Fe/H] + [$\alpha$/Fe], where [$\alpha$/Fe]
is assumed to have a constant value of 0.3 typical of Milky Way
GCs). The data, taken with the DEIMOS spectrograph on the Keck
telescope, have been restricted to GCs with S/N $\ge$12. 
We show the red and blue GCs separately. The blue GCs show
significant scatter about a mean value of [Z/H] $\sim$ --0.5, 
again suggesting a problem with their derived metallicities (see
Foster et al. 2010 for a full discussion for this and 
possible explanations). The red GCs, on the
other hand, show a clear radial gradient. Here we overplot the
total metallicity gradient derived from Figure 1 on the red GCs.
Although not a fit to the data, the metallicity gradient based on
g--i colour provides a remarkably good representation of the
overall red GC trend.

\section{Discussion}

From deep wide-field imaging we find a transition from a strong
colour gradient in both GC subpopulations which flattens to a
constant value, with the transition occurring at a projected radius of
around 6--10 arcmin. Approximately half of the GCs are 
located within this radius. 
At a slightly larger radius ($\sim$12 arcmin)
Zhang et al. (2007) report that the X-ray surface brightness
around NGC 1407 declines to a near constant value, which they
refer to as the `gas halo boundary'.

The behaviour of the NGC 1407 GC colour profiles are similar
to those in the Milky Way, for which strong gradients (--0.30 dex
per dex; Harris 2001) are
seen in both the red and blue GCs interior to 10 kpc and a
constant metallicity of [Fe/H] $\sim$ --1.5 for the blue GCs at
larger radii (there are essentially no red GCs beyond 10
kpc). 
The gradient seen in the NGC 1407 GC system is the
strongest reported to date in an elliptical galaxy, 
with the previous strongest gradient
being --0.28 dex per dex in the NGC 1399 GC system reported by
Bassino et al. (2006). We note that actual 3D gradients may be stronger than
those observed in projection. 

Most previous detections of GC colour gradients in early-type
galaxies have been  
generally restricted to $\le$ 5R$_e$ and may not extend far enough to probe
the accretion dominated region of the halos of massive galaxies.
One exception is the wide-field study of M87 by Harris (2009a)
using CFHT Megacam which allowed the GC system to be probed to
$\sim$20R$_e$. To quote Harris {\it This [metallicity gradient]
feature is clearest for R $\le$ 8 R$_e$ or about 60 kpc; beyond
that point, the red clusters are almost absent and the blue
clusters show no significant change in mean color}. Thus the GC
system of M87 also appears to reveal a transition, at a similar
scale in both effective radius and kpc 
to the NGC 1407 GC system.

The transition from a strong GC metallicity gradient to a constant
value may represent a transition from a dissipative process
(e.g. a gaseous collapse or gas-fed flows)
to a dissipationless one (e.g. accretion of low mass galaxies). 
This is in qualitative agreement with the
recent simulations of galaxy formation in two phases 
(e.g. Naab et al. 2009; Oser et al. 2010). In this model, the first {\it in-situ}
phase is associated with a dissipative collapse, which is
followed by an extended phase of ongoing accretion of low mass
galaxies. More massive galaxies have larger transition radii. 
In this picture galaxies are built from the inside-out. 
In the dissipational merger simulations of Hopkins et al. (2009)
the outer constant metallicity is determined by the metallicity
of the progenitor galaxies. 


An alternative explanation is that the observed colour profile is
caused by a changing relative mix of inner GCs, that have fixed colours, and  
outer GCs that are bluer on average. Thus as the
fraction of outer bluer GCs increases with radius a mean colour
gradient is observed until the transition radius is reached, when
the outer GCs completely dominate the overall GC colours (beyond 70 kpc in the
case of NGC 1407). Such a situation could in principle be setup
by a series of dry mergers without invoking dissipation, 
however it would need to be
fine-tuned in the sense of giving similar observed metallicity
gradients for both the blue and red subpopulations and to occur
at similar transition radii. 

A kinematic analysis of 172 GCs around NGC 1407 revealed a bias
towards tangential orbits with increasing galactocentric radius 
(Romanowsky et al. 2009). This orbital anisotropy is difficult to
reconcile with either mergers or an accretion origin for the
GCs. However, the study was largely confined to radii within 8 arcmin. 

We note that kinematic transitions at large radii have now been
reported in handful of massive 
early-type galaxies based on analysis of their stellar light and 
planetary nebulae as well as their GC systems (Proctor et al. 2009; 
Coccato et al. 2009; Foster et al. 2010; Arnold et
al. 2010). However, more work is needed to increase radial
sampling and number of galaxies studied.

Prompted by trends between the metallicity gradient of the
galaxy starlight and galaxy mass from both observations 
(e.g. Spolaor et al. 2009) and theory (e.g. Tortora et al. 2010),
we have collated the
statistically significant (i.e. more than 2$\sigma$) GC
metallicity gradients from the literature and central velocity
dispersions (a proxy for galaxy mass).
Table 1 shows that 
there is no obvious trend between the GC metallicity gradient and
the mass of the host galaxy (as traced by the central velocity
dispersion), however the sample is small and covers a limited
range in host galaxy mass and environment.

The presence of two GC subpopulations in most galaxies 
is strong evidence that galaxies did not form by a {\it simple}
monolithic collapse. The two phases of formation for NGC 1407
described here, support this view. Indeed, the {\it in-situ}
GCs appear to have formed in an early dissipative event as
envisaged by Forbes et al. (1997). Here we have shown that 
{\it both} GC subpopulations appear
to have taken part giving rise to similar strength metallicity
gradients and transition radii. The accreted GCs provide an important
contribution as envisaged in the scenario of  
Cote et al. (1998). However in 
their scenario for the GC systems of massive galaxies, 
only the red GCs were involved in a dissipative
formation event and the blue GCs were all accreted.


\begin{table*}
\caption{Globular cluster system metallicity-radius gradients in units of dex per dex. The
central velocity dispersion (in km s$^{-1}$) comes from the Hyperleda
database. The source of each globular metallicity gradient is
given in the last column. The metallicity gradients are converted
from colour gradients assuming an old stellar population. The
1$\sigma$ uncertainty when available is given in brackets.  
}
\begin{tabular}{lccccl}
\hline
Galaxy   & Environ. & $\sigma_0$ & Blue & Red & Source \\
\hline
NGC~1407 & Group & 272 & --0.38 (0.06) & --0.43 (0.07) & this work\\
NGC~4472 & Cluster & 294 & --0.15 (0.03) & --0.12 (0.06) & Geisler et al. (1996)\\
NGC~1399 & Cluster & 342 & --0.2 & --0.2 & Forte et al. (2001)\\
NGC~1399 & Cluster & 342 & --0.28 (0.06) & -- & Bassino et al. (2006)\\
M87 & Cluster & 334 & --0.12 (0.03) & --0.17 (0.03) & Harris (2009a)\\
NGC~3268 & Group & 224 & --0.11 (0.05) & -- & Harris (2009b)\\
NGC~4696 & Cluster & 254 & --0.10 (0.04) & -- & Harris (2009b)\\
NGC~7626 & Group & 271 & --0.16 (0.06) & --0.18 (0.07) & Harris (2009b)\\
6 gEs & Cluster/Group & -- & --0.10 (0.02) & --0.10 (0.02) & Harris (2009b)\\
\hline
\end{tabular}
\end{table*}



\begin{figure*}
\begin{center}
\includegraphics[angle=-90,scale=0.5]{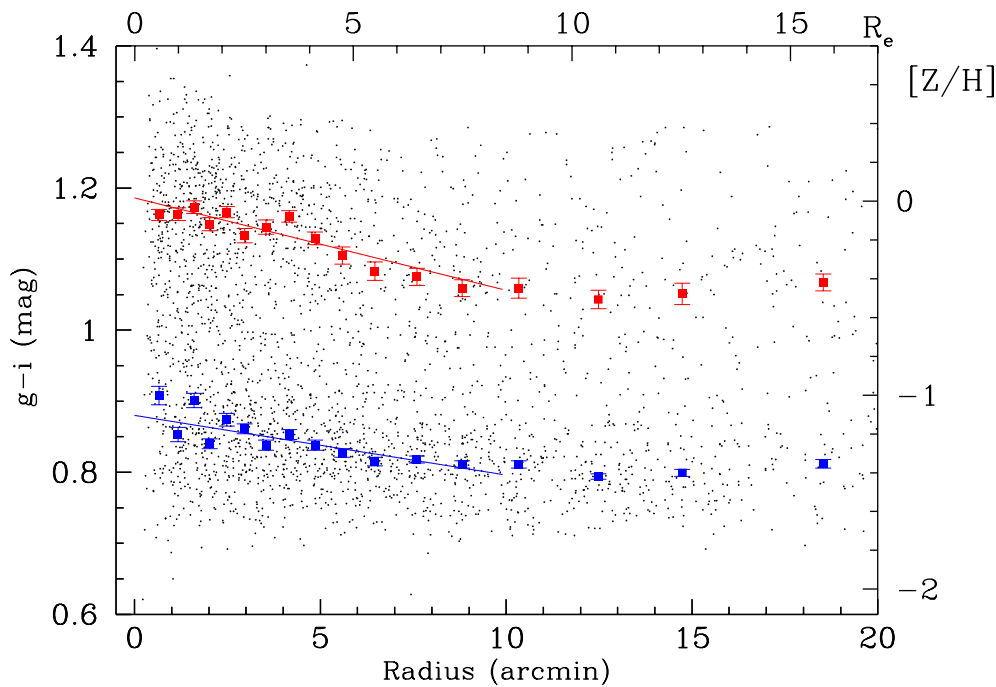}
\caption{Globular cluster colour-radial distribution from Subaru
photometry. 
The small circles
show individual GCs from Subaru photometry with i $<$ 24 (and 
errors $\le$ 0.03 mag). 
Mean colours and error on the mean are represented by
filled squares. The solid lines show the best fit linear gradient within
10 arcmin. 
The top and right axes show transformations into 
effective radius and metallicity respectively. 
Both subpopulations show a strong gradient in mean colour within the
central $\sim$10 arcmin (8.5 R$_e$, 69 kpc). The galaxy
starlight at 1 R$_e$ has a colour g--i  $\sim$ 1.2, 
similar to the mean colour of the red GCs. 
}
\end{center}
\end{figure*}


\begin{figure*}
\begin{center}
\includegraphics[angle=0,scale=0.5]{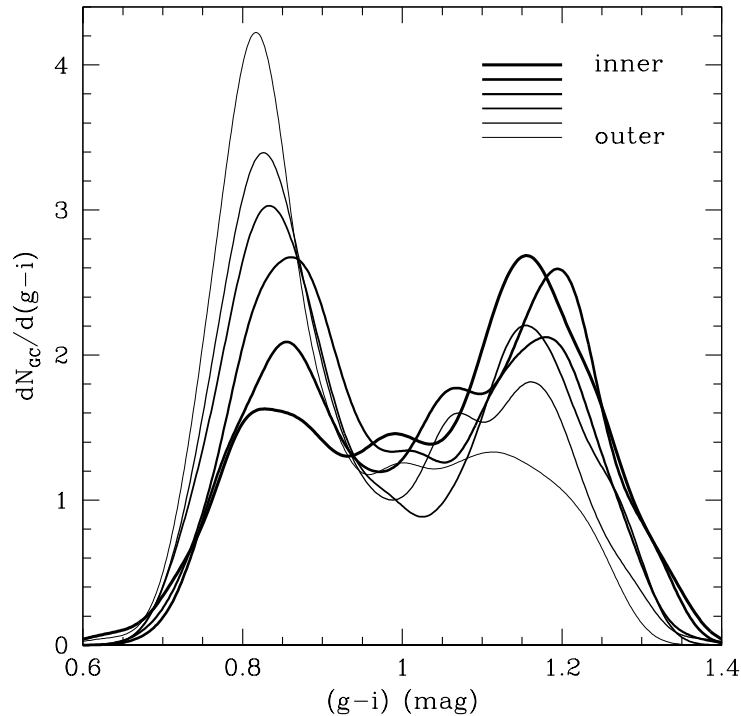}
\caption{Globular cluster colour-radius probability density
function (PDF).
The curves show the PDF in six radial bins from inner (thick
line) to outer (thin line). Other than the innermost radial bin
the blue GCs show a steady trend of getting bluer with
radius. For the red GCs the radial trend is less clear due to
possible colour substructure in the inner regions.
}
\end{center}
\end{figure*}

\begin{figure*}
\begin{center}
\includegraphics[angle=-90,scale=0.5]{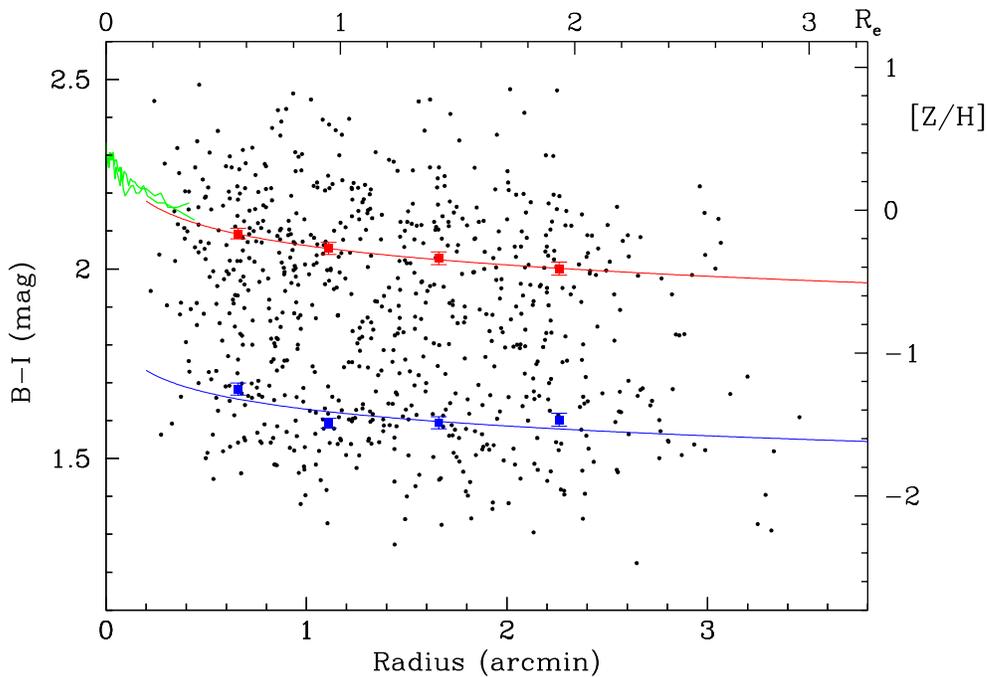}
\caption{Globular cluster colour-radial distribution from HST
photometry. The small circles
show individual GCs from HST photometry. 
Mean colours and error on the mean are represented by
filled squares. 
The top and right axes show transformations into 
effective radius and metallicity respectively. 
The solid lines show the best fit gradient in
colour-log radius space. 
The galaxy
starlight at 1 R$_e$ has a colour B--I  $\sim$ 2.2, similar to
the mean colour of the red GCs. The green lines within 0.5R$_e$
show the galaxy field star metallicity from Spolaor et
al. (2008). 
}
\end{center}
\end{figure*}

\begin{figure*}
\begin{center}
\includegraphics[angle=-90,scale=0.5]{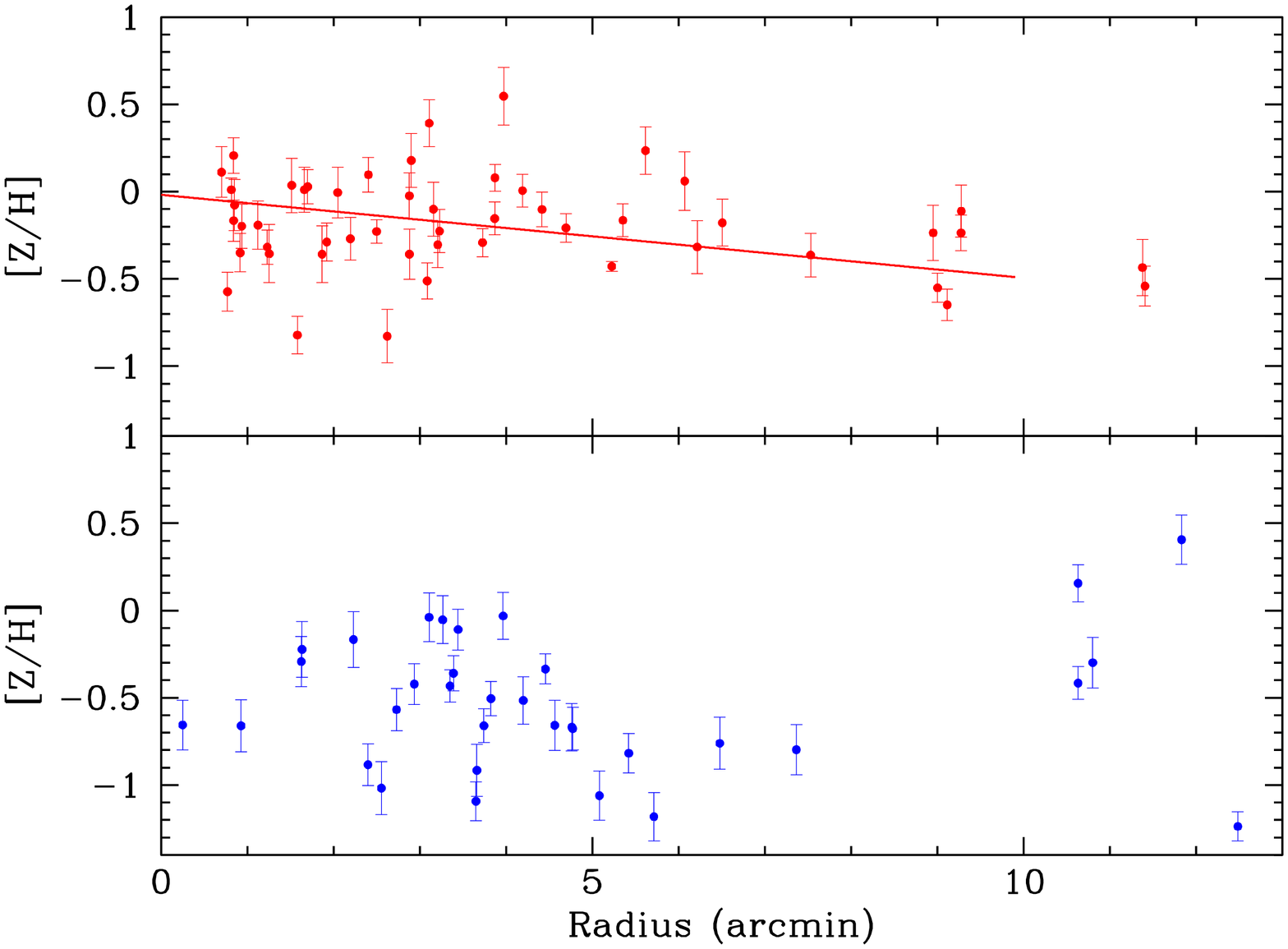}
\caption{Globular cluster metallicity-radial distribution derived
from calcium triplet measurements. 
The top panel shows the red GC subpopulation,
and the bottom panel shows the blue GC subpopulation. A clear
radial gradient is seen in the red GC subpopulation. 
The solid line shown in the top panel is not a fit to the data
but rather the metalicity gradient from Figure 1 superposed. 
}
\end{center}
\end{figure*}


\section{Conclusions}

Using deep wide-field imaging from the Subaru Suprime-cam we have
explored the variation of GC colour (and hence
metallicity) with galactocentric radius. We detect GCs out to 
$\sim$20 arcmin or 17 galaxy effective radii (R$_e$). This corresponds to
$\sim$ 140 kpc for an assumed distance of 23.77 Mpc. This is
therefore one of the most extended radial studies of a GC system
to date. 

We find that both the blue and red GC subpopulations reveal a strong
colour gradient within $\sim$6--10 arcmin (40--70 kpc) or 5--8.5 R$_e$.
Beyond this, the mean GC colours are constant. The presence of
a colour gradient within the inner 3 arcmin is supported by HST ACS imaging. 
When converted into metallicity, the gradient for the red GC
subpopulation is similar to that for the field stars in the
central $\sim$0.5R$_e$. We also report a spectroscopic metallicity gradient in
the inner 10 arcmin based on Calcium Triplet measurements from the Keck
DEIMOS spectrograph. 

The change from a strong GC metallicity gradient to a constant metallicity may
indicate the transition between an inner halo formed by {\it in-situ}
dissipative processes, with both GC subpopulations taking part,
to an outer one that is largely due
to accretion of minor mergers.  
This transition radius 
is similar to the hot gas
boundary seen in X-ray brightness profile of NGC 1407.

We find that the colour 
gradient seen in the blue GC subpopulation is not due to a changing
`blue tilt' strength with radius. Therefore, if 
the blue tilt is due to a self-enrichment process it
does not appear to have a strong radial, or environmental, dependence.



We have searched the literature for statistically significant GC
metallicity gradients in elliptical galaxies.  We do not find a
trend with host galaxy mass; however the sample is small and
largely restricted to massive galaxies in clusters/groups. We
note that previous work on the correlation between mean GC colour
and host galaxy mass has generally been restricted to galaxy
central regions. Future wide-field studies of GC system radial
colour gradients in a wider range of host galaxies and
environments may reveal new trends and hence further insight into
galaxy formation process. Such a study is relatively inexpensive
in terms of telescope time.

Our results provide observational support for the mass growth of
massive elliptical galaxies in two phases, i.e. early dissipative
collapse followed by ongoing accretion of low mass systems
(e.g. Oser et al. 2010). In other words, elliptical galaxies form 
inside-out.

\section*{Acknowledgments}

We thank V. Pota, C. Usher, C. Blom, 
and J. Arnold for their help and useful discussions. 
We also thank W. and G. Harris for their comments on a draft of
this paper. 
This project is based in part on data collected 
at the Subaru telescope, which is operated by the 
National Astronomical Observatory of Japan. Access to the Subaru
was obtained via the Gemini time swap program (GN-2006B-C-18). 
Some of the data presented herein were obtained at the W. M. Keck
Observatory, which is operated as a scientific partnership among
the Caltech, University of California and NASA. Some of the data
presented here were based on
observations made with the NASA/ESA Hubble Space Telescope.  
Finally, we thank the referee for their comments.

\end{document}